\documentclass[aps,prd,preprint]{revtex4}
\usepackage{amssymb,amsmath,graphicx}
\usepackage{graphicx}
\usepackage{mathpazo}
\newcommand{\beq}{\begin{equation}}
\newcommand{\eeq}{\end{equation}}
\newcommand{\beqa}{\begin{eqnarray}}
\newcommand{\eeqa}{\end{eqnarray}}

\begin{document}

\title{Interpretations of Quantum Mechanics: a critical survey}

\author{Michele Caponigro}

\address{Epistemology of Complexity\\ University of Bergamo, Italy \\
}

\begin{abstract}
\vspace{1cm} This brief survey analyzes the epistemological
implications about the role of observer in the interpretations of
Quantum Mechanics. As we know, the goal of most interpretations of
quantum mechanics is to avoid the apparent intrusion of the
observer into the measurement process. In the same time, there are
implicit and hidden assumptions about his role. In fact, most
interpretations taking as ontic level one of these fundamental
concepts as information, physical law and matter bring us to new
problematical questions. We think, that no interpretation of the
quantum theory can avoid this intrusion until we do not clarify
the nature of observer.
\end{abstract}
\maketitle
\baselineskip=14pt
\section{Quantum Theory: brief Overview}
\pagenumbering{arabic}\setcounter{page}{1} Can we explain what the
world is through a fundamental physical theory? This question
corresponds to the historic disagreement among scientists and
epistemologists concerning how to regard physical theories to
which people commonly refer as the realist/antirealist debate. The
position of the antirealist is the one according to which we
should not believe that physics reveals to us something about
reality but rather we should be satisfied with physics to be, for
example, just empirically adequate. In contrast, the realist is
strongly inclined to say not only that physics tells us about
reality, but also that it is our only way to actually do
metaphysics. In few words, the question is: is there an ontology?
We are interested to show through a logical pathway the existence
of a possible ontology in Nature.\\
The abstract mathematical structure of the Lorentz transformations
was deduced through simple physical principles. Thanks to the
existence of these physical principles we do not have a
significant debate on the interpretation of the theory of special
relativity. The formulation of Quantum Mechanics (QM), on to the
contrary, is based on a number of rather abstract axioms without a
clear motivation for their existence. The problem about quantum
mechanics does not lie on its effectivity, but on its
interpretation. Any attempt to interpret quantum mechanics tries
to provide a definite meaning to issues such as realism,
completeness, local realism and determinism. Despite its success,
the absence of elementary physical principles has determined a
broad discussion about the interpretation of the theory. For this
reason, and not only, Bell called the ordinary QM with the
abbreviation FAPP (for all practical purposes). The standard
interpretation of quantum mechanics, attempts, as much as
possible, to give an ontological model of physical systems using
the concept of the quantum state. However, the interpretation does
not fully succeed in giving such a model, for this reason one
solution to this problem is to abandon any attempt at an
ontological model and to put quantum mechanics on a purely
epistemological footing ( the context of informational
approaches). We believe that a possible ontological model arises
by the application of formalism of quantum mechanics to the entire
universe (including observers).\\
We will start next sections presenting, first, the basic formalism
and postulates of QM, and then overviewing some relevant
historical interpretations of QM.

\section{Postulates of quantum mechanics.}
Quantum mechanics is a mathematical model of the physical world
that describes the behavior of quantum systems. A physical model
is characterized by how it represents \emph{physical states},
\emph{observables}, \emph{measurements}, and \emph{dynamics} of
the system under consideration. A \emph{quantum system} is a
number of physical degrees of freedom in a physical object or set
of objects which is to be described quantum mechanically. The
\emph{physical state} (standard view) of a system is a
mathematical object which represents the knowledge we have about
the system, and from which all measurable physical quantities
relating to the system can be calculated. A special class of
quantum states are called the \emph{pure states}. The dimension of
$\mathcal{H}$ is a property of the degrees of freedom being
described. For example, the state of a spin-half particle lives in
a two-dimensional Hilbert space, such systems is called a quantum
bit or \emph{qubit}, and its basis vectors are labelled
$|{0}\rangle$ and $|{1}\rangle$. Pure states are sometimes called
\emph{state vectors}. We will see that more general states cannot
be described by a simple state vector, but will require a
\emph{density matrix}. The traditional way in which measurements
on quantum systems are described is in terms of
\emph{observables}. Observables are Hermitian operators which
correspond to physically measurable quantities such as energy,
momentum, spin, etc. Any Hermitian operator has a complete set of
real eigenvalues
corresponding to orthogonal eigenspaces.\\
\subsection{Basic formalism and postulates of quantum mechanics.} A
quantum description of a physical model is based
on the following concepts:\\

A \emph{state} is a complete description of a physical system.
Quantum mechanics associates a ray in \emph{Hilbert space} to the
physical state of a system.

\begin{itemize}
\item  Hilbert space is a complex linear vector space. In Dirac's ket-bra
notation states are denoted by \emph{ket vectors} $\left|
\psi\right\rangle $ in Hilbert space.

\item  Corresponding to a ket vector $\left|  \psi\right\rangle $ there is
another kind of state vector called \emph{bra vector}, which
is denoted by $\left\langle \psi\right|  $. The \emph{inner
product} of a bra $\left\langle
\psi\right|  $ and ket $\left|  \phi\right\rangle $ is defined as follows:%

\begin{align}
\left\langle \psi\right|  \left\{  \left|  \phi_{1}\right\rangle +\left|
\phi_{2}\right\rangle \right\}   &  =\left\langle \psi\mid\phi_{1}%
\right\rangle +\left\langle \psi\mid\phi_{2}\right\rangle \nonumber\\
\left\langle \psi\right|  \left\{  c\left|  \phi_{1}\right\rangle \right\}
&  =c\left\langle \psi\mid\phi_{1}\right\rangle
\end{align}

for any $c\in\mathbf{C}$, the set of complex numbers. There is
a one-to-one
correspondence between the bras and the kets. Furthermore%

\begin{align}
\left\langle \psi\mid\phi\right\rangle  &  =\left\langle \phi\mid
\psi\right\rangle ^{\ast}\nonumber\\
\left\langle \psi\mid\psi\right\rangle  &  >0\text{ for }\left|
\psi\right\rangle \neq0
\end{align}

\item  The state vectors in Hilbert space are normalized which means that the
inner product of a state vector with itself gives unity, i.e.,
\end{itemize}%

\begin{equation}
\left\langle \psi\mid\psi\right\rangle =1
\end{equation}

\begin{itemize}
\item  Operations can be performed on a ket $\left|  \psi\right\rangle $ and
transform it to another ket $\left|  \chi\right\rangle $.
There are operations on kets which are called \emph{linear
operators}, which have the following properties. For a linear
operator $\hat{\alpha}$ we have
\end{itemize}%

\begin{align}
\hat{\alpha}\left\{  \left|  \psi\right\rangle +\left|  \chi\right\rangle
\right\}   &  =\hat{\alpha}\left|  \psi\right\rangle +\hat{\alpha}\left|
\chi\right\rangle \nonumber\\
\hat{\alpha}\left\{  c\left|  \psi\right\rangle \right\}   &  =c\hat{\alpha
}\left|  \psi\right\rangle
\end{align}
for any $c\in\mathbf{C}$.

\begin{itemize}
\item  The sum and product of two linear operators $\hat{\alpha}$ and
$\hat{\beta}$ are defined as:%

\begin{align}
\left\{  \hat{\alpha}+\hat{\beta}\right\}  \left|  \psi\right\rangle  &
=\hat{\alpha}\left|  \psi\right\rangle +\hat{\beta}\left|  \psi\right\rangle
\nonumber\\
\left\{  \hat{\alpha}\hat{\beta}\right\}  \left|  \psi\right\rangle  &
=\hat{\alpha}\left\{  \hat{\beta}\left|  \psi\right\rangle \right\}
\end{align}
Generally speaking $\hat{\alpha}\hat{\beta}$ is not
necessarily equal to $\hat{\beta}\hat{\alpha}$, i.e. $\left[
\hat{\alpha},\hat{\beta}\right]  \neq0$

\item  The \emph{adjoint} $\hat{\alpha}^{\dagger}$ of an operator $\hat
{\alpha}$ is defined by the requirement:%

\begin{equation}
\left\langle \psi\mid\hat{\alpha}\chi\right\rangle =\left\langle \hat{\alpha
}^{\dagger}\psi\mid\chi\right\rangle
\end{equation}

for all kets $\left|  \psi\right\rangle $, $\left|
\chi\right\rangle $ in the Hilbert space.

\item  An operator $\hat{\alpha}$ is said to be \emph{self-adjoint} or
\emph{Hermitian} if:%

\begin{equation}
\hat{\alpha}^{\dagger}=\hat{\alpha}%
\end{equation}
\end{itemize}

Hermitian operators are the counterparts of real numbers in
operators. In quantum mechanics, the dynamical variables of
physical systems are represented by Hermitian operators. These
operators are usually called \emph{observables}.

\textbf{Postulates of quantum mechanics}:\\
Quantum theory is based on the following postulates:

\textbf{Postulate 1}: To any physical isolated system is
associated a complex vector space, where is define an inner
product (Hilbert space) which is called state space of the system.
The system is completely described by a state vector.\\
\emph{This postulate give us the universal mathematical model of
any physical system: a vector Hilbert space on the complex
numbers}\cite{Dir}.\\
\textbf{Postulate 2}: The evolution of a closed quantum system is
described by an unitary transformation. That is, the state,
%TCIMACRO{\TEXTsymbol{\vert}}%
%BeginExpansion
$\vert$%
%EndExpansion
$\psi(t)\rangle$ of the system at time $t$ is related to the state
%TCIMACRO{\TEXTsymbol{\vert}}%
%BeginExpansion
$\vert$%
%EndExpansion
$\psi(t_{0})\rangle$ a time $t_{0}$ by a unitary operator \emph{U}
which depends only on the time $t$ and $t_{0}:$
%TCIMACRO{\TEXTsymbol{\vert}}%
%BeginExpansion
$\vert$%
%EndExpansion
$\psi(t)\rangle$ = $U$
%TCIMACRO{\TEXTsymbol{\vert}}%
%BeginExpansion
$\vert$%
%EndExpansion
$\psi(t_{0})\rangle$.\\
 \emph{The second postulate describes the
temporal evolution of a closed physical system.}\\
\textbf{Postulate 3}: \emph{This postulate is about the "quantum
measurement}:
\begin{itemize}
\item  Mutually exclusive measurement outcomes correspond to orthogonal
\emph{projection operators} $\left\{  \hat{P}_{0},\text{ }\hat{P}%
_{1},...\right\}  $ and the probability of a particular
outcome $i$ is $\left\langle
\psi\mid\hat{P}_{i}\mid\psi\right\rangle $. If the outcome $i$
is attained the (normalized) quantum state after the
measurement becomes:
\end{itemize}%

\begin{equation}
\frac{\hat{P}_{i}\left|  \psi\right\rangle }{\sqrt{\left\langle \psi\mid
P_{i}\mid\psi\right\rangle}}.%
\end{equation}
Measurement made with orthogonal projection operators $\left\{  \hat{P}%
_{0},\text{ }\hat{P}_{1},...\right\}  $ is called \emph{projective
measurement}.\\
\textbf{Postulate 4}: The state space of a composite physical
system is the tensor product of the state spaces of the component
physical systems. Moreover, if we have a quantum system
$H_{i},i=1,...n$ and system $H_{i}$ is prepared in the state
%TCIMACRO{\TEXTsymbol{\vert}}%
%BeginExpansion
$\vert$%
%EndExpansion
$\psi_{i}\rangle,$ then the joint state of the total system is:
%TCIMACRO{\TEXTsymbol{\vert}}%
%BeginExpansion
$\vert$%
%EndExpansion
$\psi_{1}\rangle\otimes$\
%TCIMACRO{\TEXTsymbol{\vert}}%
%BeginExpansion
$\vert$%
%EndExpansion
$\psi_{2}\rangle$\ $\otimes.......\otimes$\
%TCIMACRO{\TEXTsymbol{\vert}}%
%BeginExpansion
$\vert$%
%EndExpansion
$\psi_{n}\rangle$\ = $H_{1}$\ $\otimes.......\otimes H_{n}.$\\
\emph{Last postulate formalizes the interaction of many physical
systems with the combination of different Hilbert spaces coming to
a unique Hilbert space.}
\begin{figure}[h]
    \begin{center}
        \scalebox{0.5}{\includegraphics*{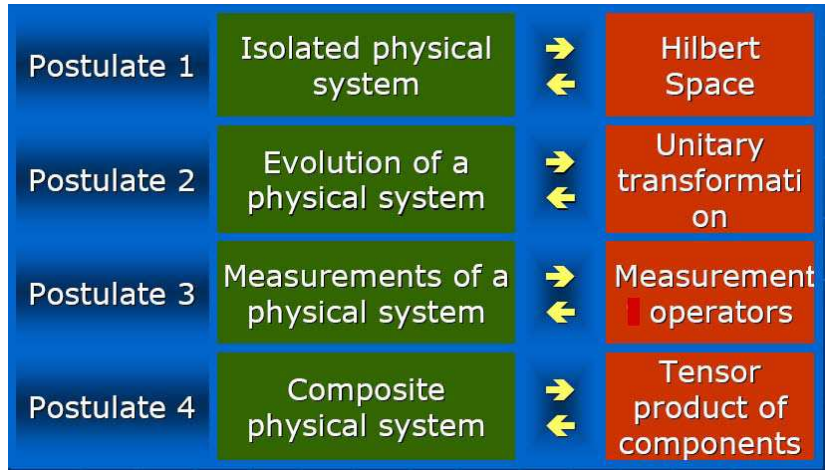}}
\caption{\label{figure} Summary of Postulates. }
\end{center}
\end{figure}

\subsection{Quantum Entanglement, Bell Inequality}
The phenomenon of quantum entanglement is widely
considered to be central to the field of quantum computation and
information. This phenomenon can be traced back to Einstein,
Podolsky and Rosen (EPR)'s famous paper \cite{EPR}\ of 1935. EPR
argued that quantum mechanical description of \emph{physical
reality} can not be considered \emph{complete} because of its
rather strange predictions about two particles that once have
interacted but now are separate from one another and do not
interact. Quantum mechanics predicts that the particles can be
\emph{entangled} even after separation. Entangled particles have
correlated properties and these correlations are at the heart of
the EPR paradox. Mathematically, the entanglement is described as
follows. For a system that can be divided into two subsystems
quantum mechanics associates two Hilbert spaces $\mathcal{H}_{A}$
and $\mathcal{H}_{B}$ to the subsystems. Assume that $\left|
i\right\rangle _{A}$ and $\left| j\right\rangle _{B}$ $($where
$i,j=1,2,...)$ are two complete orthonormal basis sets for the
Hilbert spaces $\mathcal{H}_{A}$ and $\mathcal{H}_{B}$,
respectively. The tensor product
$\mathcal{H}_{A}\otimes\mathcal{H}_{B}$ is another Hilbert space
that quantum mechanics associates with the system consisting of
the two subsystems. The tensor product states $\left|
i\right\rangle _{A}\otimes\left| j\right\rangle _{B}$ (often
written as $\left|  i\right\rangle _{A}\left| j\right\rangle
_{B}$) span the space $\mathcal{H}_{A}\otimes\mathcal{H}_{B}$. Any
state $\left|  \psi\right\rangle _{AB}$ of the composite system
made of the two subsystems is a linear combination of the product
basis states $\left|  i\right\rangle _{A}\left|  j\right\rangle
_{B}$ i.e.:%

\begin{equation}
\left|  \psi\right\rangle _{AB}=\underset{i,j}{\sum c_{ij}}\left|
i\right\rangle _{A}\left|  j\right\rangle _{B}%
\end{equation}
where $c_{ij}\in\mathbf{C}$. The normalization condition of the
state $\left| \psi\right\rangle _{AB}$ is $\sum_{i,j}\left|
c_{ij}\right|  ^{2}=1$. The state $\left|  \psi\right\rangle
_{AB}$ is called \emph{direct product }(\emph{or separable})\emph{
state} if it is possible to factor it into two
normalized states from the Hilbert spaces $\mathcal{H}_{A}$ and $\mathcal{H}%
_{B}$. Assume that $\left|  \psi^{(A)}\right\rangle
_{A}=\underset{i}{\sum }c_{i}^{(A)}\left|  i\right\rangle _{A}$
and $\left|  \psi^{(B)}\right\rangle
_{B}=\underset{j}{\sum}c_{j}^{(B)}\left|  j\right\rangle _{B}$ are
the two normalized states from $\mathcal{H}_{A}$ and
$\mathcal{H}_{B}$, respectively. The state $\left|
\psi\right\rangle _{AB}$ is a direct product state when:%

\begin{equation}
\left|  \psi\right\rangle _{AB}=\left|  \psi^{(A)}\right\rangle _{A}\left|
\psi^{(B)}\right\rangle _{B}=\left(  \sum_{i}c_{i}^{(A)}\left|  i\right\rangle
_{A}\right)  \left(  \sum_{j}c_{j}^{(B)}\left|  j\right\rangle _{B}\right)
\end{equation}
Now a state in $\mathcal{H}_{A}\otimes\mathcal{H}_{B}$ is called
\emph{entangled} if it is not a direct product state. In other
words, entanglement describes the situation when the state of
'whole' cannot be written in terms of the states of its
constituent 'parts'. Generally, it is a very hard problem to
decide whether a quantum state is entangled or not. Fortunately,
there are operational criteria, relying on measurements of
correlations, with a possible outcome from which one can conclude
that the state is entangled: the Bell inequality\cite{BELL}. A
Bell inequality is satisfied by all states which are not
entangled.Thus, if a violation of a Bell inequality is observed
the state which describes the results is entangled. Interestingly,
Bell inequalities were first introduced in a context of
\textbf{foundations of quantum mechanics}. Quantum mechanics gives
predictions in form of probabilities. Already some of the fathers
of the theory were puzzled with the question whether there can
exist a deterministic structure beyond quantum mechanics which
recovers quantum statistics as averages over "hidden variables".
In this way, it was hoped, one could get a classical-like
description which would solve the problems with the
interpretations of quantum mechanics. In his famous impossibility
proof Bell made precise assumptions about the form of a possible
underlying hidden variable structure. Spatially separated systems
and laboratories were assumed to be independent of one another
\cite{BELL}. He derived an inequality which must be satisfied by
all such (local realistic) structures. Next, he presented example
of quantum predictions which violate it. In this way the famous
Einstein-Podolsky-Rosen (EPR) paradox \cite{EPR} was solved. Bell
proved that EPR elements of reality cannot be used to describe
quantum mechanical systems.The noncommutativity of quantum theory
precludes simultaneous deterministic predictions of measurement
outcomes of complementary observables. For EPR this indicated that
"the wave function does not provide a complete description of
"physical reality". They expected the complete theory to predict
outcomes of all possible measurements, prior to and independent of
the measurement (realism), and not to allow ``spooky action at a
distance'' (locality). A more general version of Bell's theorem
for two qubits (two-level systems) was given by Clauser, Horne,
Shimony, and Holt (CHSH), and extended by Clauser and Horne (CH)
\cite{CHSH,CH}. The important feature of the CHSH and CH
inequalities, which hold for \emph{all} local realistic theories,
is that they can not only be compared with ideal quantum
predictions, but also with experimental results. The three or more
qubit versions of Bell's theorem were presented by Greenberger,
Horne, and Zeilinger (GHZ)\cite{GHZ, GHSZ}. Starting from the
assumptions of realism and locality, in 1964 Bell \cite{Bell1}
derived an inequality which was shown \cite{Aspect et al} later to
be violated by the quantum mechanical predictions for entangled
states of a composite system. As we have seen, Bell's theorem
\cite{AsherPeres} is the collective name for a family of results,
all showing the impossibility of local realistic interpretation of
quantum mechanics. Later work \cite{PeresBell} has produced many
different types of Bell-type inequalities. The Bell inequality is
expressed as follow: let $A(a)$ and $A(a^{\prime})$ be the two
observables for observer $A$ in the an EPR experiment. Similarly,
let $B(b)$ and $B(b^{\prime})$ be the two observables for the
observer $B$. In general, the observables $A(a)$ and
$A(a^{\prime})$ are incompatible and cannot be measured at the
same time, and the same holds for $B(b)$ and $B(b^{\prime})$.

It is assumed that the two particles that reach observers $A$ and
$B$ in EPR experiments\ possess hidden variables which fix the
outcome of all possible measurements. These hidden variables are
collectively represented by $\lambda $, assumed to belong to a set
$\Lambda$ with a probability density
$\rho(\lambda)$. The normalization implies:%

\begin{equation}
\int_{\Lambda}\rho(\lambda)d\lambda=1.
\end{equation}
Because a given $\lambda$ makes the four dichotomic observables
assume
definite values, we can write:%

\begin{equation}
A(a,\lambda)=\pm1;\text{ \ \ }A(a^{\prime},\lambda)=\pm1;\text{ \ \ }%
B(b,\lambda)=\pm1;\text{ \ \ }B(b^{\prime},\lambda)=\pm1
\end{equation}
That is, the physical reality is marked by the variable $\lambda$.
Now introduce a \emph{correlation function} $C(a,b)$ between two
dichotomic observables $a$ and $b$, defined by:%

\begin{equation}
C(a,b)=\int_{\Lambda}A(a,\lambda)B(b,\lambda)\rho(\lambda)d\lambda
\end{equation}
For a linear combination of four correlation functions, define
\emph{Bell's measurable quantity} $\Delta$ as:%

\begin{equation}
\Delta=C(a,b)+C(a^{\prime},b^{\prime})+C(a^{\prime},b)-C(a,b^{\prime})
\end{equation}
Only four correlation functions, out of a total of sixteen, enter
into the definition of $\Delta$. We can write:%

\begin{align}
&  \left|  C(a,b)+C(a^{\prime},b^{\prime})+C(a^{\prime},b)-C(a,b^{\prime
})\right| \nonumber\\
&  \leq\int_{\Lambda}\left\{  \left|  A(a,\lambda)\right|  \left|
B(b,\lambda)-B(b^{\prime},\lambda)\right|  +\left|  A(a^{\prime}%
,\lambda)\right|  \left|  B(b,\lambda)+B(b^{\prime},\lambda)\right|  \right\}
\rho(\lambda)d\lambda.
\end{align}
Since:%

\begin{equation}
\left|  A(a,\lambda)\right|  =\left|  A(a^{\prime},\lambda)\right|  =1
\end{equation}
we have:%

\begin{align}
&  \left|  C(a,b)+C(a^{\prime},b^{\prime})+C(a^{\prime},b)-C(a,b^{\prime
})\right| \nonumber\\
&  \leq\int_{\Lambda}\left\{  \left|  B(b,\lambda)-B(b^{\prime},\lambda
)\right|  +\left|  B(b,\lambda)+B(b^{\prime},\lambda)\right|  \right\}
\rho(\lambda)d\lambda\label{bellproof}%
\end{align}
Also $\left|  B(b,\lambda)\right|  =\left|
B(b^{\prime},\lambda)\right|  =1$, so that:%

\begin{equation}
\left|  B(b,\lambda)-B(b^{\prime},\lambda)\right|  +\left|  B(b,\lambda
)+B(b^{\prime},\lambda)\right|  =2
\end{equation}
and the inequality (\ref{bellproof}) reduces to:%

\begin{equation}
\left|  C(a,b)+C(a^{\prime},b^{\prime})+C(a^{\prime},b)-C(a,b^{\prime
})\right|  \leq2 \label{bell inequality}%
\end{equation}
which is called CHSH form \cite{CHSH}\ of Bell's inequality.
\section{\textbf{Standard Interpretation: some problems}}
Historically, the understanding of the mathematical structure of
QM went trough various stages. Very briefly, the Copenhagen
interpretation assumes two processes influencing the wavefunction,
namely, i) its unitary evolution according to the Schr\"odinger
equation, and ii) the process of measurement.

In other words, quantum  mechanics is problematic in the sense
that it is incomplete and needs the notion of a classical device
measuring quantum observables as an important ingredient of the
theory. Due to this, one accepts that there exist two worlds: the
classical one and the quantum one. In the classical world, the
measurements of classical observables are produced by classical
devices. In the framework of standard theory, the measurements of
quantum observables are produced by classical devices, too. Due to
this, the theory of quantum measurements is considered as
something very specifically different from classical measurements.

As it is well known, the Copenhagen interpretation postulates that
every measurement induces a discontinuous break in the unitary
time evolution of the state through the collapse of the total wave
function, the nature of the collapse is not at all explained, and
thus the definition of measurement remains unclear. Bohr then
followed the tenets of positivism, that implies that only
measurable questions should be discussed by scientists. Some
physicists argue that an interpretation is nothing more than a
formal equivalence between a given set of rules for processing
experimental data, thus suggesting that the whole exercise of
interpretation is unnecessary. It seems that a general consensus
has not yet been reached. Roger Penrose \cite{Penrose}, remarks
that while the theory agrees incredibly well with experiment and
while it is of profound mathematical beauty, it "makes absolute no
sense". The point of view of most physicist is rather pragmatic:
it is a physical theory with a definite mathematical background
which finds excellent agreement with experiment. So, from a
technical point of view, quantum mechanics (QM) is a set of
mathematically formulated prescriptions that deserves for
calculations of probabilities of different measurement outcomes.
The calculated probabilities agree with experiments. Pragmatic
applications of the physics are interested only in these pragmatic
aspects of QM, which is fine. Nevertheless, many physicists are
not only interested in the pragmatic aspects, but also want to
understand nature on a deeper conceptual level. Besides, a deeper
understanding of nature on the conceptual level \textbf{may} also
induce a new development of pragmatic aspects. Thus, the
conceptual understanding of physical phenomena is also an
important aspect of physics and cannot be viewed as simply
epistemological problems. The standard interpretation of QM, tells
us nothing about the underlying physics of the system. The state
vector represents our knowledge of the system, not its physics.
The main support of the standard interpretation is that
measurement process is an interaction between \emph{system and
apparatus}. This interpretation divides the world in apparatus and
system but the theory tell us nothing about these two "abstracts"
concepts. More in details, the position regarding the measurement
theory can be summarizing as following:

\begin{itemize}
    \item Measurement is an interaction between
system and apparatus.
    \item Measurements do not uncover some preexisting
physical property of a system. There is no objective property
being measured.
\item The record or result of a measurement is the only objective property.
\item Quantum mechanics is nothing more than
a set of rules to compute the outcome of physical tests to
which a system may be subjected.
\end{itemize}

This position solve most pragmatic problems but does not solve the
measurement problem, how and why occurs the collapse of the wave
function during the measurement process. The famous Schrödinger's
cat paradox is exactly this\cite{Schro}. Why the measurement
apparatus behave classically? After all it is constituted of
particles that are governed by QM rules. Where is the limit
between quantum and classical world? The following considerations
puts in evidence the problem. Consider a two-state microsystem
whose eigenfunctions are labelled by $\psi_+$ and $\psi_-$.
Furthermore, there is a macrosystem apparatus $\phi_0$, with
eigenfunctions $\phi_+$ and $\phi_-$ corresponding to an output
for the microsystem having been in the $\psi_+$ and $\psi_-$
states, respectively.  Since prior to a measurement we do not know
the state of the microsystem, it is a superposition state given by
\begin{eqnarray}
\psi_0=\alpha\psi_++\beta\psi_-,\hspace{1cm}|\alpha|^2+|\beta|^2=1.
\end{eqnarray}
Now, according to the linearity of Scr\"{o}dinger's equation, the
final state obtained after the interaction of the two systems is
\begin{eqnarray}
\Psi_0=(\alpha\psi_++\beta\psi_-)\phi_0\longrightarrow\Psi_{out}=\alpha\psi_+\phi_++\beta\psi_-\phi_-
\end{eqnarray}
where it is assumed that initially the two systems are far apart
and do not interact. The state on the far right side of the last
equation does not correspond to a definite state for a macrosystem
apparatus.  In fact, this result would say that the macroscopic
apparatus is itself in a superposition of both plus and minus
states.  Nobody has observed such macroscopic superpositions. This
is the measurement problem, since the theory predicts results that
are in clear conflict with all observations. It is \textbf{at this
point} that the standard program to resolve this problem
\emph{\textbf{invokes}} the reduction of wave packet upon
observation, that is,
\begin{eqnarray}
\alpha\psi_+\phi_++\beta\psi_-\phi_-\longrightarrow\left\{%
\begin{array}{ll}
    \psi_+\phi_+, & \hbox{$P_+=|\alpha|^2$;} \\
    \psi_-\phi_-, & \hbox{$P_-=|\beta|^2$.} \\
\end{array}%
\right.
\end{eqnarray}
Various attempts (interpretations) to find reasonable explanation
for this reduction are at the heart of the measurement problem.\\

Related to this problem, Schr\"odinger introduced his famous cat
in the very same article where entanglement was described
\cite{Schro}. Schr\"odinger devised his cat experiment in an
attempt to illustrate the incompleteness of the theory of quantum
mechanics when going from subatomic to macroscopic systems.
Schr\"odinger's legendary cat was doomed to be killed by an
automatic device triggered by the decay of a radioactive atom. He
had had trouble with his cat. He thought that it could be both
dead and alive. A strange superposition of

\begin{equation}\label{psiSchr}
|\Psi\rangle \,=\, \frac{1}{\sqrt{2}}\,\big(|\rm{excited\,atom},\rm{alive\,cat}\rangle
\,+\, |\rm{non-excited\,atom},\rm{dead\,cat}\rangle \big)
\end{equation}

\noindent was conceived. But the wavefunction (\ref{psiSchr})
showed no such commitment, superposing the probabilities. Either
the wavefunction (\ref{psiSchr}), as given by the Schr\"odinger
equation, was not everything, or it was not right. The
Schr\"odinger's cat puzzle deals with one of the most
revolutionary elements of quantum mechanics, namely, the
\textbf{\emph{superposition principle}}, mathematically founded in
the linearity of the Hilbert state space. If $|0\rangle$ and
$|1\rangle$ are two states, quantum mechanics tells us that
$a|0\rangle+b|1\rangle$ is also a possible state. Whereas such
superpositions of states have been extensively verified for
microscopic systems, the application of the formalism to
macroscopic systems appears to lead immediately to severe clashes
with our experience of the everyday world. As we have seen, the
problem is then how to reconcile the vastness of the Hilbert space
of possible states with the observation of a comparably few
"classical" macroscopic states. The long standing puzzle of the
Schr\"odinger's cat problem could be resolved in terms of quantum
decoherence. The central question of why and how our experience of
a "classical" world emerges from quantum mechanics thus lies at
the heart of the foundational problems of quantum theory.
Decoherence has been claimed to provide an explanation for this
quantum-to-classical transition. In classical physics, the
environment is usually viewed as a kind of disturbance, or noise,
that perturbs the system under consideration in such a way as to
negatively influence the study of its "objective" properties.
Therefore science has established the idealization of isolated
systems, with experimental physics aiming at eliminating any outer
sources of disturbance as much as possible in order to discover
the "true" underlying nature of the system under study. The
distinctly nonclassical phenomenon of quantum entanglement,
however, has demonstrated that the correlations between two
systems can be of fundamental importance and can lead to
properties that are not present in the individual systems. The
earlier view of phenomena arising from quantum entanglement as
"paradoxa" has generally been replaced by the recognition of
\emph{entanglement as a fundamental } property of nature. The
decoherence theory is based on the idea that such quantum
correlations are ubiquitous; that nearly every physical system
must interact in some way with its environment, which typically
consists of a large number of degrees of freedom that are hardly
ever fully controlled. Decoherence is the irreversible formation
of quantum correlations of a system with its environment. These
correlations lead to entirely new properties and behavior compared
to that shown by isolated objects, thus the decoherence seem
provides a realistic physical modelling
and a generalization of the quantum measurement process.\\
\begin{figure}[h]
    \begin{center}
        \scalebox{0.55}{\includegraphics*{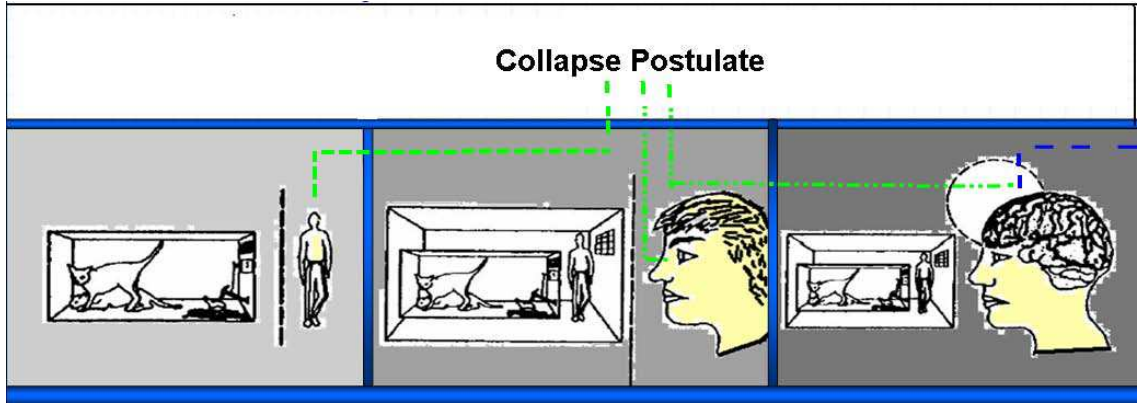}}
\caption{\label{figure} Interpretations of Collapse. }
\end{center}
\end{figure}
Next figure 2 puts in evidence the measurement problem utilizing
Schr\"{o}dinger's cat (again). The leftmost panel gives the
standard Schr\"{o}dinger cat story. There is a single observer, to
be called Ob1, outside the box. Before Ob1 opens the window to
look, the cat is in a superposition of being both alive and dead.
By opening the window and looking, Ob1 "collapses the wave-packet"
so that the cat is now in a unique state of being alive or dead.
The story gets more interesting if we place O1 in a second box as
shown in the second panel. If we, the second observer, are not
looking, then O1 is in a superposition of states seeing an alive
cat and seeing a dead cat. Once we make an observation, Ob1
collapses to one state or the other. The third panel removes the
split even further, placing it in our brain.\\
Some objections to this interpretation (standard) has been
proposed by de Muynck\cite{Mu} who fixes some fundamental points
(table and figure 3).

\begin{table}[h]
\begin{tabular}{|l|l|}
\cline{1-1}\cline{2-2}
 {\bf Positive features}   & {\bf Negative features} \\
\cline{1-1}\cline{2-2}
    +1. pragmatism     & -1. pragmatism \\
    +2. crucial role of measurement & -2. confusion of preparation
    and measurement\\
 & -3. classical account of measurement\\
 &  -4. completeness claims\\
  & -5. ambiguous notion of correspondence\\
\hline
\end{tabular}
\end{table}
According to de Muynck scheme (below), in the first realist case
a)quantum mechanics is thought to describe microscopic reality
most in the same way of classical mechanics is generally thought
to describe macroscopic reality. In the empirist case b) state
vector and density operator are thought to correspond to
preparation procedures, and quantum mechanical observables
correspond to measurement procedures and the phenomena induced by
a microscopic object in the macroscopically observable pointer of
a measuring instrument.

\begin{figure}[h]
    \begin{center}
        \scalebox{0.41}{\includegraphics*{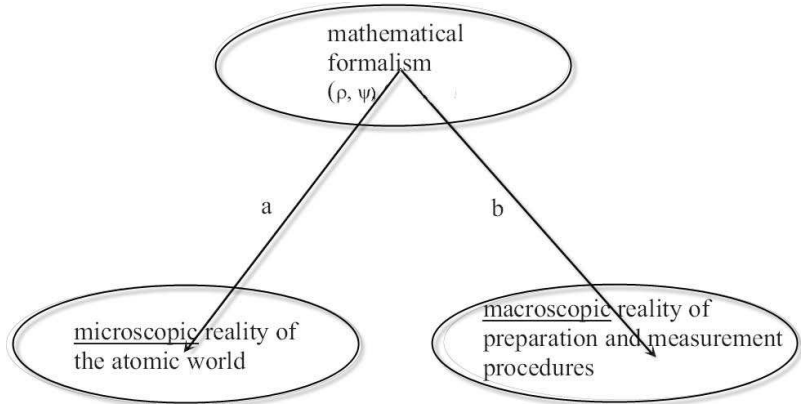}}

\caption{\label{figure}Realist (a) and empiricist (b)
interpretations of the mathematical formalism of quantum
mechanics.}
\end{center}
\end{figure}

Recently, with the development of quantum information theory,
several scientists gives to the information a fundamental role in
the description of the Nature. All these approaches start in
general from the assumption that we live in a world in which there
are certain constraints on the acquisition, representation, and
communication of information. They play on the ambiguous ontology
of quantum states. They affirm that quantum states are merely
states of knowledge (or of belief); this idea has led to the claim
that "\emph{quantum theory needs no interpretation}" \cite{Fu}.
More in details, the field of quantum information theory opened up
and expanded rapidly, for instance, quantum entanglement began to
be seen not only as a puzzle, but also as a resource which can
yield new physical effects and techniques. New insight into the
foundations of quantum physics, suggesting that information should
play an essential role in the foundations of any scientific
description of Nature. This primitive role of the information seem
to explain, according to some authors, the deep nature of physical
reality. The measurement is information \textbf{not} a physical
process. The quantum state is a construct of the observer and not
an objective property of the physical system. Some radical
positions\cite{Fu} claims that the nature of reality can be
explained as subjective knowledge. Others authors argued that
quantum theory is fundamentally just a theory of relations or of
correlations\cite{Sa}.

\section{Interpretations of QM.}
The problem linked to the collapse postulate is given in this
term: we have to consider on the one hand the temporal evolution
of the wave function \textbf{U}, provided by the rigorously
causal, deterministic and time-reversal Schr\"{o}dinger equation,
and on the other the reduction processes of the state vector, that
we call \textbf{R}. Different standpoints are possible about the
role of the processes \textbf{R} in QM. We will analyze most
important positions. We can individuate three main standpoints
about \textbf{R}:
\begin{itemize}
\item 1. The wave function contains the available information on the physical world in probabilistic
form; the wave function is not referred to an "objective
reality", but due to the intrinsically relational features of
the theory, only to what we can say about reality.
Consequently, the "collapse postulate" is simply an expression
of our peculiar knowledge of the world of quantum objects;
this is the group of \textbf{Copenhagen} and
neo-Copenhagen\cite{Mu1} interpretations.
\item 2. The wave function describes what actually
happens in the physical world
and its probabilistic nature derives from our perspective of
observers: the group of \textbf{Everett}\cite{Wa1},
Deutsch\cite{Deu},\textbf{ Bohm}\cite{She,She1} theories.
\item 3. The wave function partially describes what happens in the physical processes; in order to
comprehend its probabilistic nature and the postulate R in
particular, we need a theory connecting \textbf{U} and
\textbf{R}. This view includes all those theories which tend
to reconcile U with R by introducing new physical process:
\textbf{Penrose}\cite{Pe}, \textbf{GRW}\cite{Tu} theories.
\item 3. The wave function describes and represents an individual agent's subjective
degrees of belief. In few words, the physical reality is a
subjective information. Informational approaches
group\cite{Fu,Bu}
\\
The possible link between observer and interpretations of
Quantum Mechanics are summarized in fig.4.

\end{itemize}
\begin{figure}[h]
    \begin{center}
        \scalebox{1}{\includegraphics*{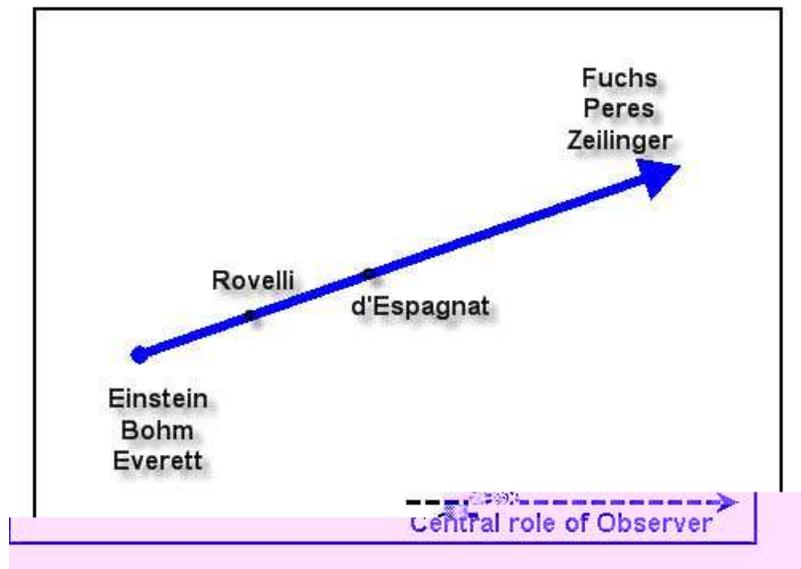}}
\caption{\label{figure} Realism to Idealism, Role of Observer. }
\end{center}
\end{figure}

\section{A possible physical reality inferred from measurement
process} We try to do a theoretical speculation on a possible
relationship between the objectivity/subjectivity nature of
measurement process and the underlying physical reality inferred.
We build the following scheme:\\

\begin{tabular}{l|l}
Measurement process   & Physical reality \\
\hline
1. ontic measurement $\longrightarrow$  & of ontic reality \\
2. ontic measurement $\longrightarrow$   & of epistemic reality \\
3. epistemic measurement $\longrightarrow$  & of ontic reality \\
4. epistemic measurement  $\longrightarrow$  & of epistemic reality\\
\end{tabular}\\

{\bf Considerations.} First case, is a realist position (without
determinism), the second, a non-completely idealistic position,
like the standard interpretation, last case is a pure idealistic
view, third position is very intriguing, we do an epistemic
measurement process but of ontic reality probably close
d'Espagnat's conception of veiled reality, a position supported
from the discovery of nonseparability in QM. According
d'Espagnat\cite{Espa} the "veiled reality" is supported from the
discovery of nonseparability in QM, he introduced the concept of
the "veiled reality" which refers to something that cannot by
studied by traditional scientific methods. d'Espagnat defines his
philosophical view as "open realism"; existence precedes
knowledge; something exists independently of us even if it cannot
be described.\\

\section{Informational Approaches to Quantum Mechanics} In
this section we introduce briefly two approaches: CBH and Fuchs'
program. All these approaches (quantum theoretic description of
physical systems) start in general from the assumption that we
live in a world in which there are certain constraints on the
acquisition, representation, and communication of information. The
concept of the information, according these approaches, play a
primary role. CBH\cite{Bu1} starting from informational
constraints try to deduce the quantum mechanics principles.
Fuchs'\cite{Fu} program involves two strong conceptual shifts: i)
quantum mechanics as a theory of information, and ii) its
probabilities as subjective degrees of belief. Utilizing the
Bayesian interpretation of probability, information assume a
subjective role. In his program, he claims that the paradoxes of
quantum mechanics, which for many interpretations provide
troubling consequences, are resolved when physical objectivity is
removed and in its place pure, subjective information is
substituted. Last, the main thesis of these approaches are
supported by the fundamentally random result of individual quantum
measurements.
\subsection{Fuchs' program: Bayesian Interpretation of Probability}
We need to analyze, how this approach interpret the notion of
probability and try to answer at fundamental questions like, what
is the nature of quantum probabilities? An ontic or epistemic
interpretation? Two agents in possession of the same facts can
assign different or the same probabilities? According this
approach we find these replies:
\begin{enumerate}
\item what is the nature of quantum probabilities?:$\Rightarrow$ \textbf{They Represent an agent's
degrees of belief.}
\item Ontic vs. Epistemic interpretation of
probabilities: $\Rightarrow$ \textbf{Epistemic
interpretation.}
\item In quantum theory, two agents in possession of the same facts can assign
different or the same probabilities?:$\Rightarrow$
\textbf{Different probabilities}
\item It is indispensable for the description of physical reality to introduce the agent?$\Rightarrow$ \textbf{Yes}
\end{enumerate}
The central role played by Bayes theorem is learning from
experimental data. The theorem teaches how the probability of each
hypothesis has to be updated in the light of the new observation.
For instance, to solve a problem via Bayes' theorem mean: to know
the outcome of a series of observations of the system and to want
to estimate its properties (state, parameters). The Bayesian
interpretation of quantum mechanics is founded on the notion that
quantum states, both pure and mixed, represent states of knowledge
and that all the probabilities they predict are Bayesian
probabilities.\\
There are many objections, for instance: how we choice the priors
(subjective priors)to enter in the bayesian inference? Priors are
pointed to by those critical of the Bayesian approach as the major
weakness of the theory.

\section{Conclusion}
As we have seen, every interpretation, in a different ways, claims
to explain the "observer" and the underlying physical reality once
established as ontic level, one of three fundamentals elements:
information, matter or physical law. We have presented some
problems related these affirmations. We think, that no
interpretation of the quantum theory can avoid this intrusion
until we do not clarify the nature of observer.\\

{\footnotesize------------------\\ $\diamond$\emph{Michele Caponigro}\\
University of Bergamo
\\
$\diamond$ \emph{michele.caponigro@unibg.it} }

\end{document}